\documentclass{sigchi-ext}
\usepackage[T1]{fontenc}
\usepackage{textcomp}
\usepackage[scaled=.92]{helvet} %
\usepackage{graphicx} %
\usepackage{balance}  %
\usepackage{booktabs} %
\usepackage{ccicons}  %
\usepackage{ragged2e} %
\usepackage{amsmath}
\usepackage{tasks}
\usepackage{csquotes}
\usepackage{color,soul}

\def\plaintitle{Helpful, Misleading or Confusing: How Humans Perceive Fundamental Building Blocks of Artificial Intelligence Explanations}

\def\emptyauthor{Edward Small, Yueqing Xuan, Danula Hettiachchi, Kacper Sokol}
\def\plainkeywords{Evaluation; validation; human-centred; user study; artificial intelligence; machine learning; explainability; interpretability.}

\title%
{Helpful, Misleading or Confusing: \\
How Humans Perceive Fundamental Building Blocks of Artificial Intelligence Explanations}%

\numberofauthors{4}
\author{%
    \textbf{Edward Small}\textsuperscript{\textdagger},
    \textbf{Yueqing Xuan}\textsuperscript{\textdagger}, \\
    \textbf{Danula Hettiachchi}\textsuperscript{\textdagger}, and
    \textbf{Kacper Sokol}\textsuperscript{\textdagger} \\
    \affaddr{ARC Centre of Excellence for Automated Decision-Making and Society, School of Computing Technologies, RMIT University, Australia} \\
    \email{edward.small@student.rmit.edu.au} \\
    \email{yueqing.xuan@student.rmit.edu.au} \\
    \email{danula.hettiachchi@rmit.edu.au} \\
    \email{kacper.sokol@rmit.edu.au}%
}

\definecolor{linkColor}{RGB}{6,125,233}
\hypersetup{%
  pdftitle={\plaintitle},
  pdfauthor={\emptyauthor},
  pdfkeywords={\plainkeywords},
  bookmarksnumbered,
  pdfstartview={FitH},
  colorlinks,
  citecolor=black,
  filecolor=black,
  linkcolor=black,
  urlcolor=linkColor,
  breaklinks=true,
}

\usepackage{todonotes}
\usepackage[justification=justified]{caption}
\usepackage{subcaption}
\usepackage{newfloat}
\usepackage{eurosym}

\DeclareFloatingEnvironment[fileext=frm,placement={!ht},name=Question]{question}
\newenvironment{marginquestion}[1][-1.2ex]%
  {\begin{@tufte@margin@float}[#1]{question}}
  {\end{@tufte@margin@float}}
  
\DeclareFloatingEnvironment[fileext=frm,placement={!ht},name=Explanation]{explanation}
\newenvironment{marginexplanation}[1][-1.2ex]%
  {\begin{@tufte@margin@float}[#1]{explanation}}
  {\end{@tufte@margin@float}}

\newcommand{\answer}[1]{\textcolor{red!75}{\hl{#1}}}

\begin{document}

\copyrightinfo{\\\emph{2023 ACM CHI Workshop on Human-Centered Explainable AI (HCXAI)}\\[0.5\baselineskip]\textsuperscript{\textdagger}~All authors contributed equally to this work.}

\maketitle

\justifying
\sloppy

\begin{abstract}
Explainable artificial intelligence techniques are developed at breakneck speed, but suitable evaluation approaches lag behind. %
With explainers becoming increasingly complex and a lack of consensus on how to assess their utility, it is challenging to judge the benefit and effectiveness of different explanations. %
To address this gap, we take a step back from sophisticated predictive algorithms and instead look into explainability of simple decision-making models. %
In this setting, we aim to assess how people perceive comprehensibility of their different representations such as mathematical formulation, graphical representation and textual summarisation (of varying complexity and scope). %
This allows us to capture how diverse stakeholders -- engineers, researchers, consumers, regulators and the like -- judge intelligibility of fundamental concepts that more elaborate artificial intelligence explanations are built from. %
This position paper charts our approach to establishing appropriate evaluation methodology as well as a conceptual and practical framework to facilitate setting up and executing relevant user studies.%
\end{abstract}


\section{Motivation}

It is often challenging to envisage how novel explainable artificial intelligence (XAI) and interpretable machine learning (IML) techniques will be perceived by explainees given their diverse skills, experiences and background knowledge~\cite{sokol2021explainability}. %
The gap between design intention and actual reception of explanatory insights may sometimes be larger than expected -- thus compromising their utility -- especially when serving a lay audience~\cite{ehsan2021explainable,keenan2023mind}. %
To alleviate such problems and enable better design and operationalisation of explainability systems, we should strive to understand the unique needs of diverse audiences as well as their specific interpretation of and satisfaction with different explanation types. %

The availability of tools to peek inside opaque predictive models has ballooned in recent years~\cite{dovsilovic2018explainable}, nevertheless it remains unclear what constitutes a ``good'' explanation~\cite{sokol2021explainability}. %
To complicate matters further, a good explanation for one user may be unintelligible for another~\cite{kaur2020interpreting,sokol2021explainability}. %
Similarly, certain explanations may be easy to misinterpret or misunderstand~\cite{ribeiro2016should}, leading some users to overly rely on or misplace their trust in these insights~\cite{schrills2020color} (akin to \emph{negative comprehension}~\cite{kaur2020interpreting}). %
This can result in harmful explainers that appear faithful and trustworthy enough to convince a user of their truthfulness and utility while at the same time offering hollow or factually incorrect information~\cite{kruger1999unskilled}. %

Such scenarios motivate us to assess the intelligibility of \emph{fundamental building blocks} found across XAI and IML tools, which tend to be complex and multifaceted sociotechnical systems tested predominantly as end-to-end explainers~\cite{sokol2022and}. %
By explicitly accounting for how different demographics perceive distinct types and presentation modalities of explanations, our user study aspires to highlight the necessity of a comprehensive human-centred design and evaluation framework in this research space. %

\section{Objective}%

We aim to examine how users from diverse backgrounds perceive explanations of automated decision-making (ADM) systems. %
In particular, we intend to explore whether explanatory insights foster \emph{warranted} (and justified) or \emph{unwarranted} trust in predictive models~\cite{ferrario2022explainability,jacovi2021formalizing}. %
To this end, we propose three measures suitable for dedicated (online) user studies, hereafter referred to as the \textit{3-C evaluation framework}. %
\begin{description}\compresslist
    \item [Comprehension] The gap between the information offered by an algorithmic explanation and the information understood by an explainee (i.e., new knowledge).%
    \begin{description}[font=\it]%
        \item [Positive] comprehension captures \emph{known knowns} -- the user knows what information an explanation offers -- and \emph{known unknowns} -- the user knows what information it lacks.%
        \item [Negative] comprehension encompasses \emph{unknown knowns} -- the user is unable to interpret the information communicated by an explanation -- and \emph{unknown unknowns} -- the user is ignorant of the limitations of an explanation.%
    \end{description}
    \item [Confidence] The degree of an explainee's certainty in their own understanding of an explanation.%
    \item [Contentment] The explainee's perception of how reliable and informative an explanation is.%
\end{description}

\section{Methodology}\label{method}

To ensure the simplicity of our study as well as generalisability of our findings, we investigate a comprehensive and representative subset of predictive models, explanation types and modalities thereof. %
We focus exclusively on explanations that are derived directly from the tested models to guarantee their full fidelity and truthfulness~\cite{rudin2019stop}. %
Specifically, we draw prototypical data-driven predictive models from across their distinct categories: linear, quadratic and the like from the \emph{geometric} family; logistic regression from the \emph{probabilistic} group; and (shallow) decision tree from the \emph{logical} collection~\cite{flach2012machine}. %
We look into \emph{algorithmic explanations} that explicitly communicate the model's (mathematical) operation as well as %
\emph{observational explanations} that capture diverse aspects of its predictive behaviour through different categories -- \emph{associations between antecedent and consequent}, \emph{contrasts and differences}, and \emph{causal mechanisms} -- and types -- \emph{model-based}, \emph{feature-based}, and \emph{instance-based} -- of explanatory insights~\cite{sokol2020explainability}. %
Notably, this broad range of explanations accounts for their varied scope (local, cohort or global) and limitations (whether implicit or explicit). %
We present them through different media -- (statistical or numerical) summarisation, visualisation, textualisation, formal argumentation, and a mixture thereof -- testing both \emph{static} and \emph{interactive} modalities~\cite{sokol2020one,sokol2018glass}.%

\begin{marginfigure}[-2\baselineskip]%

  \begin{minipage}{.9999\marginparwidth}
    \centering
  \begin{minipage}{.95\textwidth}
  $$
  f(\mathbf{x}) = 0.7 x_1 - 0.3 x_2
  $$
  \end{minipage}
  \caption{Mathematical representation of a model.\label{fig:poly_model}}%
  \end{minipage}

  \vspace{1pc}

  \begin{minipage}{.9999\marginparwidth}
    \centering
    \includegraphics[width=.9\textwidth]{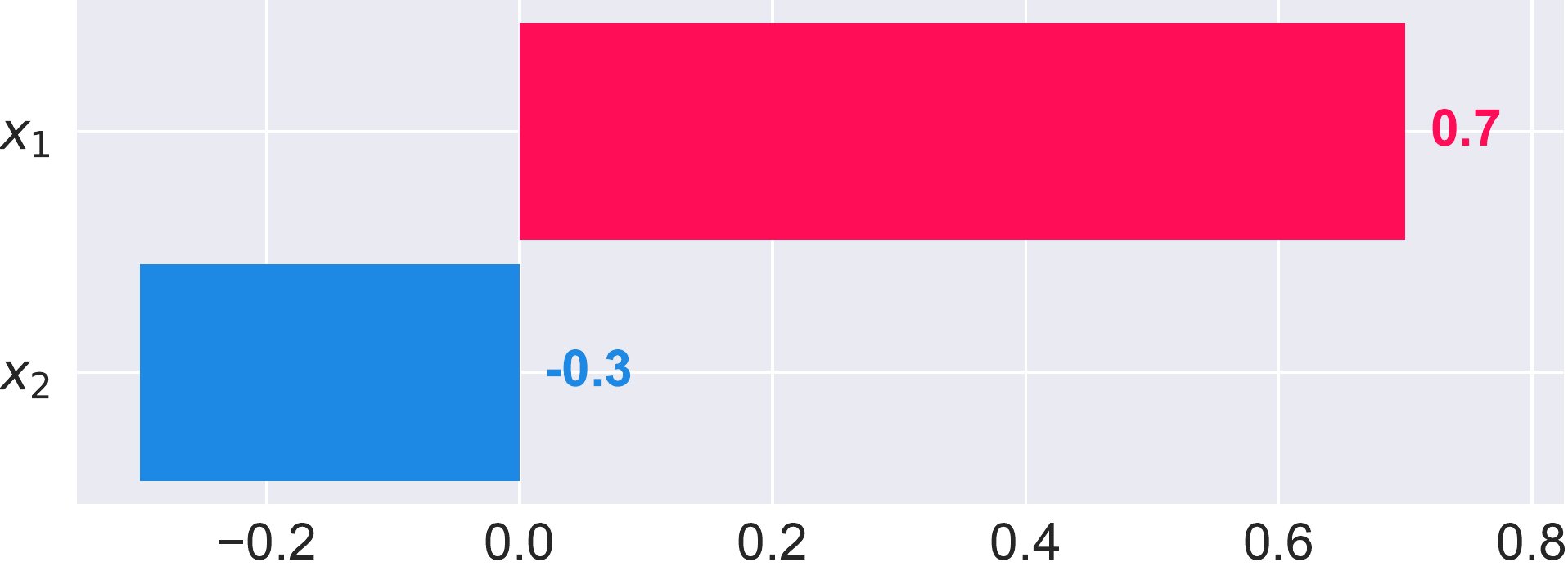}%
    \caption{Visualisation of feature influence/importance derived from a model (e.g., its parameters).\label{fig:feature_importance}}%
  \end{minipage}

  \vspace{1pc}

  \begin{minipage}{0.98\marginparwidth}
    \centering
    \includegraphics[width=.7\textwidth]{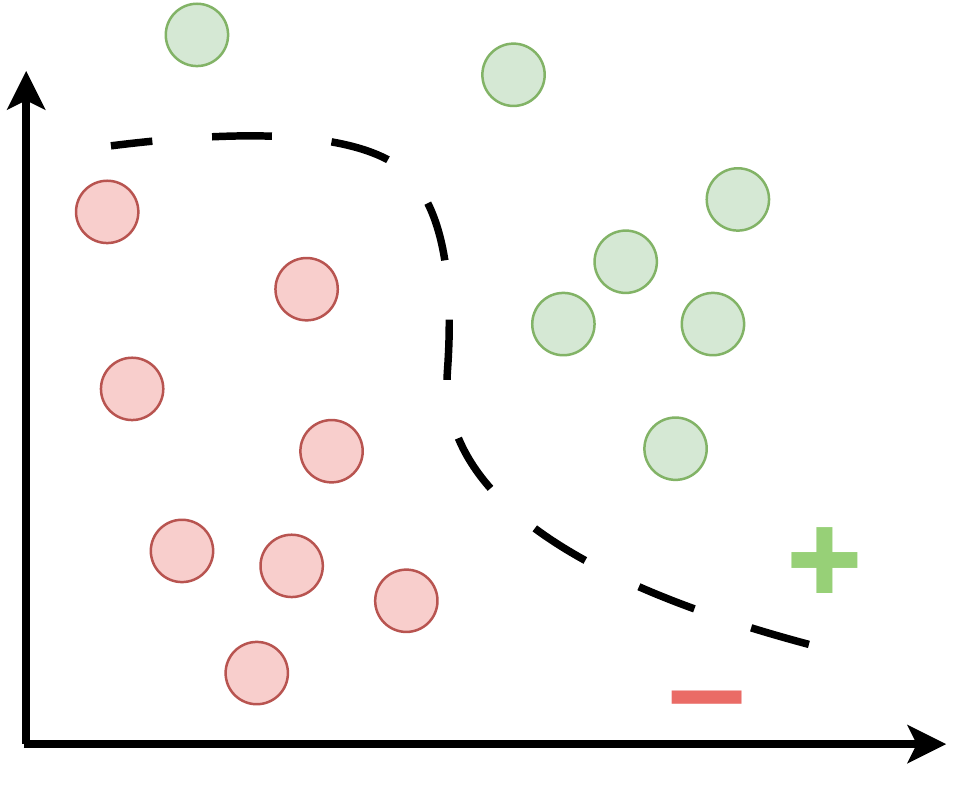}%
    \caption{Visualisation of a decision boundary used by a predictive model.\label{fig:boundary}}%
  \end{minipage}

  \vspace{1pc}

  \begin{minipage}{0.98\marginparwidth}
    \centering
    \includegraphics[width=.35\textwidth]{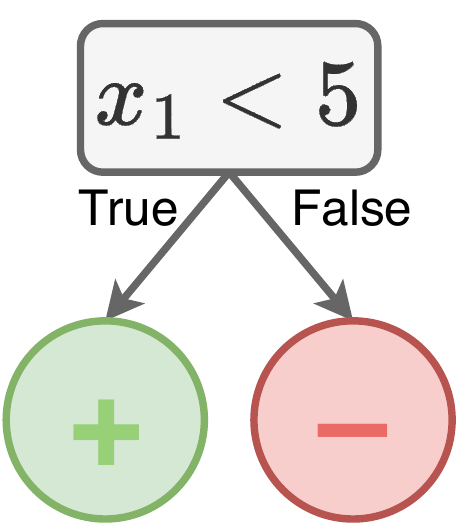}%
    \caption{Visualisation of a decision tree structure.\label{fig:tree}}%
  \end{minipage}

  \vspace{1pc}
  
  \begin{minipage}{.9999\marginparwidth}
    \centering
  \begin{minipage}{.75\textwidth}
  \it
      Had your income been \euro{}10,000 more,\\you would have been awarded the loan.
  \end{minipage}
  \caption{Textual counterfactual explanation.\label{fig:cf}}%
  \end{minipage}
\end{marginfigure}

For example, a geometrical model can be explained via %
its mathematical formulation (Figure~\ref{fig:poly_model}), %
its parameter values (Figure~\ref{fig:feature_importance}), %
the visualisation of its decision surface within the desired space (Figure~\ref{fig:boundary}), %
exemplars sourced from both sides of its decision boundary, or what-if statements. %
Similarly, a logical model, e.g., a decision tree, can be visualised as a diagram (Figure~\ref{fig:tree}) and explained with decision space partition, logical rules (root-to-leaf paths), feature importance, exemplars extracted from leaves as well as what-ifs and counterfactuals (Figure~\ref{fig:cf}) derived directly from the model~\cite{sokol2020limetree}.%

To better understand how the perception of different explanations varies across the participants, we envisage collecting basic demographic information such as age, gender, educational attainment and profession. %
Literacy is another important dimension that affects explanation comprehensibility~\cite{ehsan2021explainable}; %
while it is difficult to capture in a generic setting, we will adapt pre-existing scales~\cite{van2016development} to assess explainees' \textit{digital literacy} (ability to interact with technology and familiarity with artificial intelligence techniques), \textit{English proficiency} and \textit{numeracy} (ability to interpret mathematical concepts). %

The explainees' needs as well as the function, depth and scope of explanations must also be considered since XAI and IML tools are used by diverse stakeholders to different ends, e.g., engineers (technical expertise), consumers (no expertise) and auditors (limited technical knowledge)~\cite{belle2021principles,laato2022explain}. %
Among these audiences, the general public may care only about the logic leading to a particular decision; %
regulators may need to access and assess the end-to-end functioning of a predictive system; and %
practitioners may require the same insight but with more technical depth. %
Evaluating the utility of explanations in view of varied needs, expectations and expertise can yield important findings that will benefit the designers and consumers of XAI and IML systems~\cite{meske2022explainable}. %

The within-subject online user study will be based on \emph{linear}, \emph{polynomial}, \emph{logistic} and \emph{decision tree} models, and a diverse set of explanations spanning model presentation (Figures~\ref{fig:poly_model} \& \ref{fig:tree}), model summarisation (Figure~\ref{fig:feature_importance}), decision boundary visualisation (Figure~\ref{fig:boundary}) and textual counterfactuals (Figure~\ref{fig:cf}). %
We will rely on familiar and unfamiliar domains (see the next section) to simulate background knowledge and lack thereof, e.g., loan application~\cite{van2021effect} and medical diagnosis~\cite{cai2019hello}. %
For every predictive model and data domain, the participants will complete three tasks, each based on two explanations; %
in the process, they will answer targeted questions and provide feedback on the explanatory artefacts. %
The study workflow for each task is envisaged as follows: %
\begin{enumerate}\compresslist
    \item Present an automated decision. %
    \item Offer its explanation using method A (Explanation~\ref{ex:loan}). %
    \item Measure comprehension (Questions~\ref{q:one} \& \ref{q:three}) and confidence (Question~\ref{q:two}). %
    \item Measure contentment (Question~\ref{q:four}) with user rating of the quantity and quality, communication, and reliability of the information provided by the explanation. %
    \item Offer another explanation based on method B. %
    \item Re-evaluate the 3-Cs. %
\end{enumerate}

\section{Case Studies}

\begin{marginexplanation}[.5\baselineskip]%

  \vspace{-7.0\baselineskip}

  \begin{minipage}{0.98\marginparwidth}
    \centering
    \caption{Loan application.}%
    \textit{The loan request was rejected because of your income (\euro{}42,000).\label{ex:loan}}
  \end{minipage}
\end{marginexplanation}


\begin{marginquestion}%
  \begin{minipage}{0.98\marginparwidth}
    \centering
    \caption{Comprehension.\label{q:one}}%
    \textit{Income was the strongest factor in this specific loan application.}\\[.5\baselineskip]%
    \begin{tasks}(2)
    \task \answer{True.}
    \task False.
    \task Can't say.%
    \task Don't know.%
    \end{tasks}
  \end{minipage}

  \vspace{1pc}
  
  \begin{minipage}{0.98\marginparwidth}
    \centering
    \caption{Confidence.\label{q:two}}
    \textit{How confident are you in your previous answer?}\\[.5\baselineskip]
    \begin{tabular}{p{2cm}|p{2cm}|p{2cm}|p{2cm}|p{2cm}|p{2cm}|p{2cm}|}
    \hline
    \multicolumn{1}{|c|}{1} & \multicolumn{1}{c|}{\answer{2}} & \multicolumn{1}{c|}{3} & \multicolumn{1}{c|}{4} & \multicolumn{1}{c|}{5} & \multicolumn{1}{c|}{6} & \multicolumn{1}{c|}{7} \\ \hline                     
    \end{tabular}
  \end{minipage}
  
  \vspace{1pc}

  \begin{minipage}{0.98\marginparwidth}
    \centering
    \caption{Comprehension.\label{q:three}}%
    \textit{Increasing the \textit{monthly loan repayment} to \euro{}400 will change the automated decision.}\\[.5\baselineskip]
    \begin{tasks}(2)
    \task True.
    \task False.
    \task \answer{Can't say.}%
    \task Don't know.%
    \end{tasks}
  \end{minipage}

  \vspace{1pc}
  
  \begin{minipage}{0.98\marginparwidth}
    \centering
    \caption{Contentment.\label{q:four}}
    \textit{The explanation was easy to interpret.}\\[.5\baselineskip]
    \begin{tabular}{p{2cm}|p{2cm}|p{2cm}|p{2cm}|p{2cm}|p{2cm}|p{2cm}|}
    \hline
    \multicolumn{1}{|c|}{1} & \multicolumn{1}{c|}{\answer{2}} & \multicolumn{1}{c|}{3} & \multicolumn{1}{c|}{4} & \multicolumn{1}{c|}{5} & \multicolumn{1}{c|}{6} & \multicolumn{1}{c|}{7} \\ \hline                     
    \end{tabular}
  \end{minipage}

\end{marginquestion}

To ensure the \emph{ecological validity} of our study, i.e., the generalisability of its findings, we will consider \emph{real} and \emph{fictitious} scenarios spanning \emph{low-} and \emph{high-stakes} domains. %
Because the individual perception of each scenario's impact can be subjective~\cite{gillespie2023trust}, we will ensure that one case study is always recognised as having less at stake than the other, albeit low-stakes cases may not necessarily be of low impact to all the users across the board~\cite{crockett2020risk}. %
When designing specific scenarios, we will draw inspiration from relevant application domains listed in the General Data Protection Regulation (GDPR) to ensure that they are relatable and align with real-life ADM systems~\cite{voigt2017eu}. %
Such a comprehensive approach will allow us to better understand the 3-Cs of XAI and IML tools in a variety of settings. %

When designing the low- and high-stakes ADM case studies, we will either %
embed them in reality~\cite{leichtmann2023effects,robbemond2022understanding} or %
narrate them in a fictitious context that is clearly disconnected from the real life~\cite{narayanan2018humans}. %
An example of a real-world, high-stakes domain is a financial trading assistant powered by an ADM system; %
in this case the explainees risk incurring monetary loss if they misplace their trust in or misunderstand an insight output by the agent. %
Alternative high-stakes scenarios can be situated in a justice (e.g., pretrial bail), healthcare (e.g., administration of a life-altering medication) or job screening contexts. %
A real-world, low-stakes domain, on the other hand, may be based on using ADM tools to identify bird species or assist in playing board games. %
Similar case studies can be embedded in unfamiliar contexts by asking explainees to: %
follow recommendations of an ADM medical decision support tool to decide on a treatment of a sick extraterrestrial (high-stakes), and %
vet outfits recommended to an extraterrestrial by an ADM personal stylist based on attire preferences and body type (low-stakes). %

Both fictitious and real-world use cases come with pros and cons. %
By employing the former we can prevent the explainee's background knowledge and preferences from influencing their interaction with XAI and IML systems~\cite{narayanan2018humans}. %
For example, user study participants cannot use their pre-existing medical expertise when dealing with a fictitious extraterrestrial healthcare problem, which forces them to rely exclusively on the information provided by an explanation. %
However, such a setting can compromise the ecological validity of our study since fictitious domains may be perceived as abstract and unrelatable, decreasing the participants' motivation and engagement. %
Despite the extraterrestrial healthcare setting being inherently high-stakes, user study participants may be unable to judge the severity and consequences of misunderstanding ADM systems given the unrealistic narrative. %
This phenomenon may be difficult to capture and implicitly impact our 3-Cs when comparing them between low- and high-stakes domains since the latter setting may not be perceived as such by some participants. %

Real-world use cases are inevitably affected by the background knowledge of user study participants, which varies from person to person, is difficult to capture or account for, and may bias the measurement of our 3-Cs. %
Such a setting, however, ensures ecological validity of the study findings and maximises (active) engagement of the participants (especially for high-stakes domains) since they are on the receiving end of ADM systems~\cite{leichtmann2023effects}. %
Additionally, the influence of background knowledge can be reduced by substituting relevant terminology with made-up nomenclature, enabling our 3-C framework to capture the desired properties. %
While we will explore both real and fictitious settings in our preliminary study, as it stands we lean towards real-world scenarios. %

\section{Discussion}

The target audience of XAI and IML systems is at the heart of effective, human-understandable explanations of automated decisions. %
Each group of people has different requirements, preferences and expectations with respect to the explanations, fulfilling which secures their trust and confidence in the delivered information. %
In view of this, %
we envision to capture how the demographic characteristics of explainees may affect explanation comprehensibility and elucidate the anticipated variability through the lens of our 3-C framework. %
This will help us to analyse what types of explanations are preferred by laypersons and which modalities impose usability barriers. %

In particular, we are interested in the manifestation of information overload -- a situation in which explainees fail to process incoming information as it overwhelms their cognitive abilities. %
The trade-off between richness and utility of information varies to different degrees by explanation type and modality; %
for example, providing exhaustive textual explanations is likely to fail in communicating high-impact information, which in such scenarios can be easily overlooked. %
Exploring the acceptance threshold is therefore critical in advancing XAI and IML human-centred evaluation frameworks. %

Explaining the overall operation of an automated decision-making system is just as important as doing so for an individual output. %
While the outcome of an ``opaque'' predictive model can be justified, such an explanation is fundamentally different from an inherently transparent model that is open to scrutiny~\cite{lipton2018mythos,rudin2019stop}. %
Within this purview, is there a point where transparency -- allowing an agent to access the raw, unadulterated model -- does not improve understanding~\cite{sokol2021explainability}? %
We anticipate to answer this question with our 3-C framework. %

Ensuring that explanations are equally interpretable across distinct demographics and populations is a formative step in advancing fairness of data-driven systems~\cite{mehrabi2021survey}. %
By collecting basic demographic information about the user study participants, we can assess if individuals from different groups perceive certain explanations as equally comprehensible and trustworthy~\cite{van2021effect}. %
Algorithmic bias is another concern. %
Predictive models are generally developed and trained in Global North~\cite{siddarth2021ai}, which may result in implicit biases and prevent these systems from generalising to underrepresented demographics who differ in how they handle cognitive information. %
Common considerations include circumstances such as a discrepancy in literacy and ability to interpret textual or visual semantics~\cite{holmes1997women,van2021effect}. %

\section{Conclusion and Future Work}

In this position paper, we set out to assess how differences in explainees' literacy and education attainment, among many other traits, shape their interpretation of algorithmic explanations. %
We developed a quantitative framework to measure these aspects from three distinct perspectives: comprehension, confidence and contentment. %
Our 3-C evaluation paradigm offers comprehensive and nuanced insights into how fundamental building blocks of XAI and IML systems are interpreted. %
The proposed approach is a first step towards a grounded evaluation methodology based on user studies that accounts for individual differences between explainees. %

In future work, we will focus on quantitative assessment of explainee literacy by designing structured knowledge tests. %
We also plan to use real-life explainers across different domains, such as loan approval (high-impact) and animal identification (low-impact), spanning tabular (e.g., generic numerical and categorical characteristics) and sensory (e.g., specialised medical imaging) data. %

\section{Acknowledgement}
This research was conducted by the ARC Centre of Excellence for Automated Decision-Making and Society (project number CE200100005), and funded by the Australian Government through the Australian Research Council.%

\balance{}

\bibliographystyle{SIGCHI-Reference-Format}
\bibliography{sample-base}

\end{document}